\documentclass[prb,showpacs]{revtex4}
\usepackage{amsmath,amssymb,graphicx,bezier}
\usepackage{comment}
\newcommand{\nc}{\newcommand}
\nc{\rnc}{\renewcommand}
\nc{\nn}{\nonumber}
\nc{\ch}{\cosh}
\nc{\sh}{\sinh}
\nc{\sech}{{\rm sech}}
\nc{\bra}{\langle}
\nc{\ket}{\rangle}
\rnc{\H}{\mathcal{H}}
\nc{\M}{\mathcal{M}}
\rnc{\(}{\left(}
\rnc{\)}{\right)}
\rnc{\Im}{{\rm{Im}\,}}
\rnc{\Re}{{\rm{Re}\,}}
\nc{\tr}{{\rm Tr}}
\nc{\Lam}{\Lambda}
\nc{\ep}{\varepsilon}
\nc{\calH}{{\cal H}}
\nc{\calA}{{\cal A}}
\nc{\calL}{{\cal L}}
\def\bm#1{\mbox{\boldmath $#1$}}
\begin{document}
%
\title
{
Universal temperature dependence of the magnetization of gapped spin chains
}
\author{Yoshitaka Maeda}

\affiliation{ 
The University of British Columbia, 
Vancouver, British Columbia, Canada V6T 1Z1}

\author{Chisa Hotta}
\affiliation{Aoyama Gakuin University, 229-8558, Sagamihara, Japan}

\author{Masaki Oshikawa}

\affiliation{ Institute of Solid State Physics, University of Tokyo,
Kashiwa 277-8581, Japan}

%
\begin{abstract}
Temperature dependence of the magnetization of the Haldane spin chain
at finite magnetic field is analyzed systematically. 
Quantum Monte Carlo data indicates a clear minimum of magnetization 
as a function of temperature in the gapless regime. 
On the basis of the Tomonaga-Luttinger liquid theory, 
we argue that this minimum is rather universal and can be observed 
for general axially symmetric quasi-one-dimensional spin systems. 
Our argument is confirmed by the magnetic-field dependence of 
the spin-wave velocity obtained numerically. 
%
One can estimate a magnitude of the gap of any such systems 
by fitting the experimental data with the magnetization minimum. 
\end{abstract}
\maketitle
\section{Introduction}\label{sec:intro}

The discovery of the Haldane gap~\cite{Haldane} established 
an important paradigm followed by the extensive studies 
on gapped spin chains. 
In particular, recent remarkable progress in high magnetic field
experiments provide various data on the closing of the gap
by the applied magnetic field.
They stimulate a renewed interest in corresponding theoretical studies. 
However, because of methodological reasons 
the theories on gapped spin systems still lie far behind 
those of the exactly solvable $S=1/2$ Heisenberg chain~\cite{QISM}. 
Actually at present the only clue to understand the experiments 
are the phenomenological effective theories~\cite{Affleck,Tsvelik} 
and the numerical calculations which support them. 
\par
In this paper, we study the temperature dependence of 
the magnetization of the $S=1$ Haldane chain
for wide range of magnetic field. 
We particularly focus on the cases where the magnetic field, 
$h$, exceeds the gap, $\Delta$. 
The magnetization is one of the most fundamental
quantity observed in experiments.
In spite of the simplicity of the problem, it is not easy to give a systematic understanding of the magnetization in such systems. Numerical approaches to the problem also had difficulties.
For example, quantum Monte Carlo (QMC) method~\cite{Suzumasu,Roji,LUA} 
had difficulty 
due to small acceptance ratio in a finite magnetic field. 
This was overcome very recently\cite{Pro1,Pro2,SSE1,SSE2,DL} with
new formulations such as
Stochastic Series Expansion (SSE).
Taking advantage of such developments, we revisit the problem by 
combining various analytical approaches with
numerical results obtained with modern techniques
(QMC with SSE, and density matrix renormalization group(DMRG)).
We clarify universal features
in the temperature dependence of the magnetization in
the Haldane chain under an applied magnetic field.
Our findings are applicable to general gapped
one-dimensional spin systems, as well.

\par

The Hamiltonian of the Heisenberg spin chain is given by 
\begin{align}\label{eq:Heisenberg}
\H = J\sum_{j=1}^N \bm S_j \cdot \bm S_{j+1}-h\sum_{j=1}^N S_j^z,
\end{align}
where $J>0$ and $N$ is the system size. 
The ground state of $S=1$ Haldane chain is non-magnetic (a singlet state) 
and has a finite energy gap, $\Delta=0.4105J$, which 
is estimated numerically~\cite{White, Todo}. 
The system undergoes a quantum phase transition at finite magnetic field, 
$h=h_c(=\Delta)$; 
in the Haldane gap phase $h<h_c$, the magnetization 
exponentially vanishes toward zero temperature, 
which is confirmed experimentally, e.g. in Ref.~\onlinecite{NENP}. 
On the other hand, in the $h\ge h_c$ phase (gapless regime) 
the magnetization is finite even at $T=0$. 
This regime can be described by the Tomonaga-Luttinger (TL) 
liquid\cite{Affleck} as in the $S=1/2$ Heisenberg chain. 
Therefore, one might expect that its magnetization also 
follows that of the $S=1/2$ Heisenberg chain, 
which is well-known as the Bonner-Fisher curve~\cite{BF}. 
This is, however, not the case. 
We show that the magnetization curve as a function of $T$ 
in the $S=1$ gapless regime has a minimum at $T=T_m$ and that 
the value of $T_m$ depends on the magnetic field. 
Although this behavior resembles that of the three-dimensional Bose-Einstein
condensation (BEC) of magnons~\cite{NOOT}, 
the origin obviously differs between the two cases 
since the BEC at $T\ne0$ never occur in the one-dimensional systems~\cite{BEC}.  
We prove  
that this phenomena can be observed in general one-dimensional spin systems 
(chains and ladders) with the
rotational symmetry about the magnetic field direction (axial symmetry).  
\par
The magnetization of the $S=1$ gapless phase has already been studied in 
many articles; 
the field dependence of the magnetization is presented 
at the several fixed values of temperature~\cite{Okunishi,LouQin,Kashu}, 
while the particular details of its temperature dependence is overlooked. 
Hida {\it et al.} discussed in the sine-Gordon model~\cite{Hida} the winding numbers induced by 
the chemical potentials, 
which can be interpreted as the magnetization due to the finite magnetic field 
in the Haldane chain. 
This mapping describes correctly the $h\sim h_c$ case.
 However, it breaks down at higher magnetic fields. 
Konik {\it et al.}~\cite{KonikFendley} derived the susceptibility $\chi(h)$ 
as a function of $h$ on the basis of 
the exact solution of the nonlinear sigma (NL$\sigma$) model 
(see $M=\int \chi(h)dh$ in Fig.5 of Ref~\onlinecite{KonikFendley}). 
Unfortunately, at finite magnetization, the NL$\sigma$ model 
does not give accurate descriptions as we will discuss shortly. 
In this way, the comprehensive understanding of magnetization in the entire gapless regime 
$h_c<h<h_s$ was lacking, which we clarify in the present paper. 
Various universal features are found which also apply to other 
one-dimensional gapped spin systems. 
\par
We present our findings in the following manner; In Sec.~\ref{sec:numerical} 
we give the QMC results for the magnetization of the $S=1$ Haldane chain.
In Sec.~\ref{sec:dis}, 
we elucidate the universal features of the temperature dependence of the
magnetization, and explain the numerical results.
Our findings on purely one-dimensional systems are also compared with
the magnon Bose-Einstein condensation in three dimensions.

\section{Numerical Results}\label{sec:numerical}
\par
\begin{figure}[htb]
\begin{center}
\includegraphics[width=7cm]{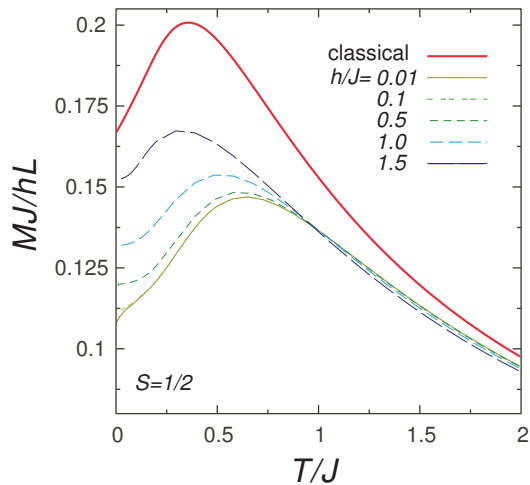}
\end{center}
\caption{The temperature dependence of the magnetizations of the $S=1/2$
 Heisenberg model. The exact classical spin solution and the Bethe
 ansatz results for several choices of ($h/J=0.01 \sim 1.5$) are depicted.}
\label{fig:half}
\end{figure}

Let us first remind of the temperature dependence of the 
magnetization of the $S=1/2$ Heisenberg chain 
as a reference. 
At present precise numerical data are easily obtained on the 
basis of the Bethe Ansatz solution 
(using the integrability of the Quantum Transfer Matrix~\cite{QTM})
as shown in fig.~\ref{fig:half} 
for the various magnetic fields. 
For any value of the magnetic field, as a function of the temperature, the magnetization monotonically increases until it hits a maximum, and then monotonically decreases down to zero in the infinite temperature limit.
This behavior is qualitatively the same as in the classical spin result~\cite{Fisher}. 
The XXZ chain in the gapless regime also shows a similar behavior. 
The decrease at low temperature is intuitively interpreted as the growth 
of the antiferromagnetic short range order.
\par
\begin{figure}[htb]
\begin{center}
\includegraphics[width=7cm]{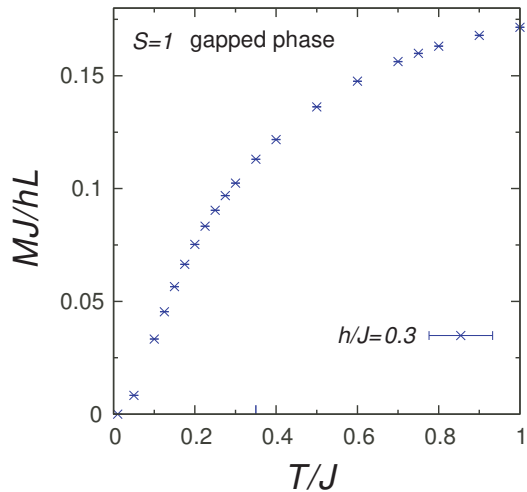}
\end{center}
\caption{The temperature dependence of the magnetization of the $S=1$
 Heisenberg model for $h(<h_c)$: $h=0.3J$. In this regime, the
 temperature dependence monotonically decreases;
 $T^{-1}\exp(-\Delta/T)$, especially for $T\ll\Delta$.
The error bar of each point is much smaller than the symbol size.}
\label{fig:lowh}
\end{figure}
\begin{figure}[htb]
\begin{center}
\includegraphics[width=14.5cm]{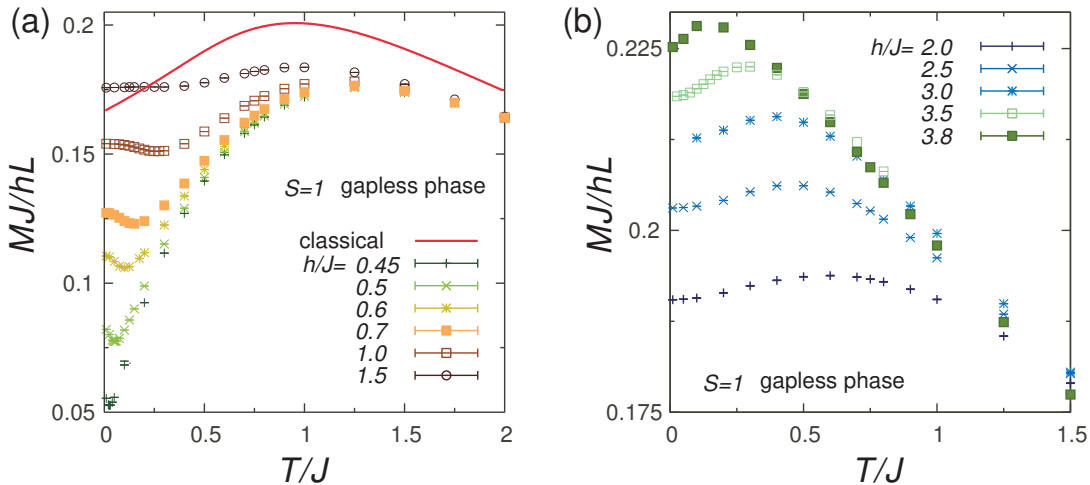}
\end{center}
\caption{The temperature dependence of the magnetization of the $S=1$
 Heisenberg model in the gapless phase($h>h_c$) below the 
saturation field($h_s=4.0J$) 
at several values of $h/J$, (a)0.45-1.5, and (b)2.0-3.8. 
The exact susceptibility of the classical spin Hamiltonian is shown 
together in (a). 
The error bar of each point is much smaller than the symbol size.}
\label{fig:magh}
\end{figure}

Next, we present the QMC results for the $S=1$ Haldane chain. 
The QMC simulation is performed at $N=512$ with the maximum $1200000$ 
Monte-Carlo step using a stochastic series expansion~\cite{SSE1,SSE2,DL} 
code of the ALPS library~\cite{ALPS,Alet,Troyer}. 
The convergence in the thermodynamic limit after the finite size scaling 
is confirmed by the comparison of our results with the high temperature expansion results. 
Figures~\ref{fig:lowh} shows the magnetization in the Haldane gap phase $h<h_c$
of eq.~(\ref{eq:Heisenberg}). 
As well known, it monotonically decreases and vanishes toward $T\to 0$ 
following $T^{-1}\exp(-\Delta/T)$. 
In contrast, the magnetization at $h>h_c$ has a characteristic structure. 
Figures~\ref{fig:magh}(a) and \ref{fig:magh}(b) show the magnetizations
at $h_c<h<h_s=4J$ ($h_s$ is the saturation field). 
As in the $S=1/2$ case, the qualitative behavior is consistent with that of the classical spin 
system except at low temperature.
However, unlike the $S=1/2$ case the magnetization minimum is observed at $T=T_m$. The temperature $T_m$, which gives the magnetization minimum, decreases toward $T=0$ as $h$ is lowered to $h_c$.
%
\section{Discussion}\label{sec:dis}
\par
We discuss the characteristic behavior found in Fig.~\ref{fig:magh} 
by use of generic effective theories of the gapped spin chains with short-range interactions. 
Note that the logic given in this section are not restricted to the
Haldane chain. 
\subsection{Effective theory near $h_c$ and $h_s$}
\begin{figure}[htb]
\begin{center}
\includegraphics[width=8cm]{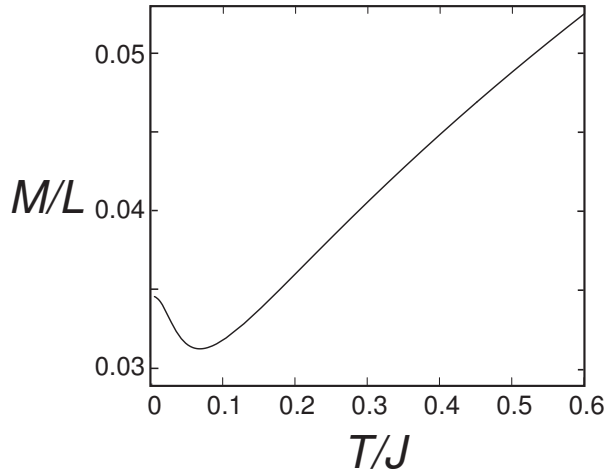}
\end{center}
\caption{The free fermion result, eq.~(\ref{eq:polylog}) for $h-\Delta=0.3$, is depicted.}
\label{fig:fermi}
\end{figure}

Excited states above the gap from the $S^z=0$ ground state 
generally consists of a triplet massive boson state with $S^z=\pm 1,0$ 
which usually have repulsive short-range interactions. 
Low-energy states of these particles may be approximated by a non-relativistic dispersion relation: 
\begin{align}\label{eq:dis2}
E(k)\approx
\Delta+k^2/2m-hS^z.
\end{align}  
where 
$m$ is a band curvature. 
If an underlying relativistic theory exists, which is not necessarily
the case,  $v_0=\sqrt{\Delta/m}$ corresponds to the relativistic ``speed of light''. 
As a result, the $S^z=1$ magnon branch intersect with the ground state at 
$h = h_c=\Delta$, a quantum phase transition point. 
In the gapless regime a quasi long range order appears~\cite{Affleck}, 
where one may integrate out the $S^z=0,-1$ magnons which are the higher energy states. 
Then, in the low density limit of the magnons, $h= h_c$, the remaining 
$S^z=1$ magnons can be exactly mapped onto the free fermion theory with the dispersion 
 eq.~(\ref{eq:dis2})~\cite{Lieb,Affleck,Schulz}. 
The low but finite magnon density regime, $h\gtrsim h_c$, can be still
approximately described by the free fermion theory 
since residual interactions are proportional to $h-h_c$. 
Here, the magnetization is equal to the number of the particle as 
\begin{align}\label{eq:fermi}
\frac{M}{L}
=
\sqrt{\frac{m}{2\pi^2}}
\int^{\infty}_0 d\epsilon 
D(\epsilon) f(\epsilon-\mu), 
\end{align}
where $\epsilon=k^2/2m$, $\mu=h-\Delta$, and $f(\epsilon-\mu)=1/\{\exp(\beta(\epsilon-\mu))+1\}$ 
is the Fermi distribution function. 
The density of states follow, $D(\epsilon)\propto 1/\sqrt{\epsilon}$, constant, 
and $\sqrt{\epsilon}$ in one, two, and three-dimensions, respectively. 
Then, from eq.~(\ref{eq:fermi}), 
the increase of the magnetization at $T\to0$
in Fig~\ref{fig:magh} can be easily understood by the following argument; 
the Fermi distribution function is symmetric with respect to the Fermi
point $\mu$, namely,
$f(\epsilon+\delta\epsilon-\mu)=1-f(\epsilon-\delta\epsilon-\mu)$. Moreover,
$f(\epsilon-\mu)$ for $\epsilon>2\mu$ near $T=0$ 
is negligibly small, since
$f(\epsilon-\mu)\sim \exp(-\beta(\epsilon-\mu))$.
 Therefore, if the density of states 
increases for lower energy $\epsilon$, so does eq.~(\ref{eq:fermi}) at $T\to0$. 
From this observation, we can conclude that this decrease of the
magnetization in fig.~\ref{fig:magh} stems from the singularity 
of the density of states in one-dimensional systems.
The exact integration of eq.~(\ref{eq:fermi}) gives 
the explicit expression, 
\begin{align}\label{eq:polylog}
\frac{M}{L}
=-\sqrt{\frac{m}{2\pi\beta}}{\rm Li}_{n=1/2}[-e^{\beta (h-\Delta)}], 
\end{align}
where
${\rm Li}_n[x]=\sum_{l=1}^\infty
x^l/l^n$ is the polylogarithm function. 
Actually, the numerical description of eq.~(\ref{eq:polylog}) in fig.~\ref{fig:fermi} 
reproduces well the minimum of $M$ found in fig.~\ref{fig:magh}(a).  
\par
The analogous mapping between the dilute boson and the free fermion 
holds at $h \lesssim h_s$ as well. 
In this case, the vacuum state is the fully-polarized state 
and the low-lying excitations consist of $S^z = 0$ magnon branch 
instead of the $S^z=1$ one. 
Then, the magnetization shows a maximum at low temperature 
corresponding to the minimum of the magnon density, 
just the opposite to that at $h\gtrsim h_c$. 
The behavior found in fig.~\ref{fig:magh}(b) (e.g $h=3.5, 3.8$) are 
consistent with this argument. 
\par
%
\subsection{Effective theory in the low-energy limit for $h_c\le h\le h_s$}
The free fermion description has thus succeeded in reproducing the 
minimum/maximum in fig.~\ref{fig:magh}. 
However, it is valid only slightly above $h_c$ and below $h_s$. 
On the other hand, the TL liquid theory should be applicable for 
the whole gapless regime in one-dimensional systems 
($h_c < h < h_s$ in the present case),
albeit only in the low energy limit. 
In the following, we apply the TL liquid theory to estimate 
the magnetization. 
From the conformal field theory, the low temperature 
expansion of the free energy per unit volume 
is given by~\cite{AffleckCFT,Cardy}  
\begin{align}\label{eq:freeenergy}
f=\epsilon_0-\frac{\pi c}{6v_F}T^2+O(T^3),
\end{align}
where $\epsilon_0$ is an ground state energy, 
$v_F$ is the excitation velocity of a fixed point theory, 
and $c$ is the central charge($c=1$ for TL liquid). 
The magnetization is given by the derivative of the free
energy with respect to the magnetic field,  
\begin{align}\label{eq:TLexpa}
\frac{M}{L}
&=-\partial_h f\nonumber\\
&=\frac{M_0}{L}-\frac{\pi}{6v_F^2}\frac{\partial v_F}{\partial h} T^2
+O(T^3).
\end{align}
The first and the second term give the magnetization of the ground state 
and the leading finite temperature correction, respectively. 
This equation indicates that whether 
the magnetization near $T=0$ increases or decreases 
is determined by the sign of the gradient of the velocity 
with respect to the magnetic field, $\partial_h v_F$. 
For the $S=1/2$ Heisenberg chain, we always have 
$\partial_h v<0$ (see fig.~9 in ref.~\onlinecite{AO}), 
which is consistent with Fig.~\ref{fig:half}. 
In contrast, the gapless TL liquid regime in the present 
case has $v_F=0$ at both endpoints, $h=h_c$ and $h=h_s$, 
so that $v_F$ must have a maximum in between. 
This indeed gives the characteristic 
behavior observed in Fig.~\ref{fig:magh}. 
\par
Let us demonstrate the validity of eq.~(\ref{eq:TLexpa}) in the special case of $h \gtrsim h_c$, where an explicit evaluation of $v_F(h)$ is available. 
Following the free-fermion dispersion in eq.~(\ref{eq:dis2}) 
we have 
\begin{align}\label{eq:vfree}
v_F=v_0
\sqrt{
\frac{2(h-\Delta)}{\Delta}
}.
\end{align}
The low temperature expansion of the magnetization given by the free-fermion theory is derived from 
the Sommerfeld expansion~\cite{AshcroftMermin} on eq.~(\ref{eq:fermi}) as, 
\begin{align}\label{eq:expansion}
\frac{M}{L}=
\frac{\sqrt{2\Delta(h-\Delta)}}{\pi v_0}
-\frac{\pi}{12v_0}\sqrt{\frac\Delta2}(h-\Delta)^{-3/2}T^2
+O(T^4), 
\end{align}
with $v_0^2=\Delta/m$. 
These two equations lead to 
\begin{align}\label{eq:mexpafermi}
\frac{M}{L}
=
\frac{\sqrt{2\Delta(h-\Delta)}}{\pi v_0}
-
\frac{\pi}{6v_F^2}\frac{dv_F}{dh}T^2+O(T^4),
\end{align}
which is the special case $h\gtrsim h_c$ of the general result,
eq.~(\ref{eq:TLexpa}). 
The behavior of $M/L$ at $h\lesssim h_s$ is analogously obtained by 
replacing $v_F$ in eq.(\ref{eq:vfree}) with $v_F\propto \sqrt{h_s-h}$, 
following the mapping of $S^z=1$ at $h\gtrsim h_c$ magnons with $S^z=0$ 
magnons at $h\lesssim h_s$. 
\par
To confirm the above argument we obtain $v_F$, by 
the DMRG calculation 
on the $S=1$ Haldane chain with periodic boundary condition. 
The energy at the fixed magnetization, $M/L$, is obtained for discrete values of 
$M=0-L$ with system size up to $L=120$. 
One can fit the energy values, $E(M/L,L)$, obtained after 
the extrapolation of truncation error to the finite size scaling:
\begin{align}\label{eq:dmrg}
\frac{E(M/L,L)}{L}= \epsilon_o(M/L)- \frac{\pi v_F(M/L)}{6}\frac{1}{L^2} 
+O(1/L^3), 
\end{align}
where $\epsilon_o(M/L)$ and $v_F(M/L)$ denote the energy and velocity at 
finite magnetization, respectively. 
\begin{figure}[htb]
\begin{center}
\includegraphics[width=13cm]{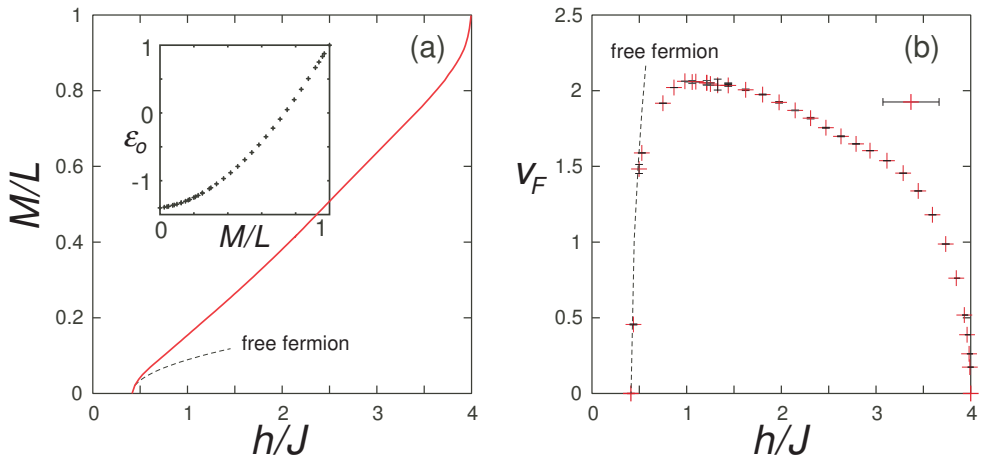}
\end{center}
\caption{(a) Magnetization as a function of magnetic field 
of the $S=1$ Haldane chain in the bulk limit. 
The inset shows the $M/L$ dependence of energy per site 
obtained by DMRG with periodic boundary condition up to $L=120$. 
The energy at each $L$ is extrapolated against the truncation error, 
where the accuracy is guaranteed to less than the order of $10^{-6}$ 
in the worst case. 
(b)The magnetic field dependence of velocity derived from the data 
in (a). 
The broken lines in (a) and (b) are the free fermion values taken 
from the first term of eq.(9) and from eq.(7) with $v_0/J=2.49$~\cite{Sorensen}, respectively.}
\label{fig:dmrg}
\end{figure}
Figure~\ref{fig:dmrg}(a) shows the magnetization curve which is 
obtained from the energy as a function of $M/L$ in its the inset. 
By use of this $M/L-h$ curve, we plot in Fig.~\ref{fig:dmrg}(b) 
the velocity as a function of the magnetic field. 
The velocity increases rapidly just above $h=h_c$, takes a 
maximum at around $h\sim J$, and then decreases toward $h=h_s$, as 
anticipated in the previous discussion. 
The reason why the gradient of $v_F$ shows a sudden change at 
$h\sim J$ and $3J$ is still out of our scheme, while the shape of $v_F$ 
in the whole gapless regime suggests that it might be divided into 
three parts according to the change of nature in its excitation 
spectrum. 
Going back to Fig.~\ref{fig:magh} 
we find several cases where the $M-T$ curves are almost flat in a fairly 
wide temperature range, e.g., $h/J=1.5,2.0$. 
Apparently, this is consistent with the fact that 
the velocity $v_F$ takes the maximum ($\partial_h v_F=0$)
at a certain field within this range, as shown in Fig.~\ref{fig:dmrg}(b). 
At this field, the leading finite temperature correction 
starts from $O(T^3)$. 
Even when $h$ is slightly shifted from the maximum, $v_F$ depends rather weakly on $h$. Thus for the range of $h$ around the maximum, magnetization looks almost flat in the low temperature regime.
\par
\par 
So far, we have not discussed the effects of irrelevant operators on
eq.~(\ref{eq:freeenergy}). 
Generally, Lorentz invariant terms never renormalize 
the spin-wave velocity which corresponds to the ``speed of light''. 
Therefore, to examine whether the corrections due to irrelevant operators 
are required in eq.~(\ref{eq:TLexpa}), 
perturbations which break Lorentz invariance should be 
explicitly considered.  
In principle, irrelevant perturbations could give larger corrections 
than $T^2$, 
e.g. in the case of the $S=1/2$ Heisenberg chain, 
the marginally irrelevant operator gives a logarithmic correction 
to the zero magnetic field susceptibility~\cite{EAT}. 
Generally, under the compactification of the bosonic field $\phi$, 
the vertex operator $\exp{(i\alpha\phi)}$ ($\alpha$ is a constant parameter) 
and $(\partial_x^m\phi)^n$ (m and n are integers) can be perturbations 
around the TL liquid fixed point. 
In the present case with $M\neq 0$, the former one 
became less relevant in the low temperature regime, $T\ll M/J$. 
This is because it usually takes a form of 
$\exp(iMx/L)\times \exp(\alpha\phi)$, remaining negligible in the low
energy limit. 
Then, the $(\partial_x\phi)^3$-term becomes the leading perturbation, 
since the $\partial _x\phi$ and $(\partial_x\phi)^2$-terms are exactly absorbed 
into $\calL_0$ and $\partial_x^2\phi$ is forbidden 
by the parity symmetry ($x\to-x$ and $\phi\to-\phi$). 
By a standard dimension analysis, 
this term again leads to a $T^2$ correction to eq.~(\ref{eq:TLexpa})
. This correction should be understood as already included in eq.~(\ref{eq:TLexpa}), which thus stands valid.

\subsection{Crossover temperature and estimation of the gap}
\begin{figure}[htb]
\begin{center}
\includegraphics[width=7.5cm]{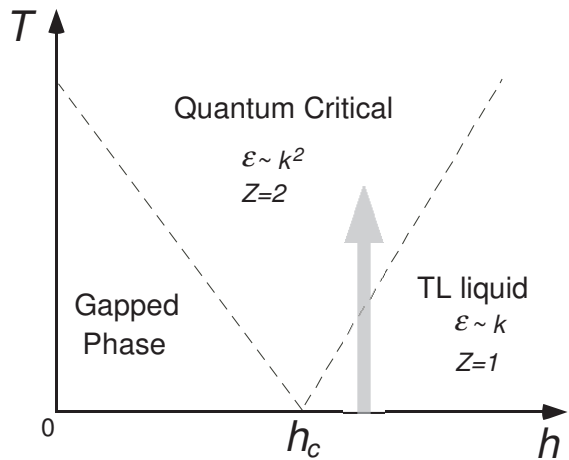}
\end{center}
\caption{The phase diagram for the gapped spin chains. Dashed lines
 indicate the crossovers.
The crossover temperature between the TLL regime and the quantum
 critical regime is $T\sim h-h_c$.}
\label{fig:crossover}
\end{figure}

Let us now return to the discussion to the characteristic 
temperature that give the magnetization minimum, $T_m$.
It can be interpreted as a crossover temperature from the TL liquid 
to the state with non-relativistic dispersion, $\epsilon\propto k^2$, 
which is indicated by the arrow in fig.~\ref{fig:crossover}. 
At $T\ll h-h_c$, the system can be described by a TL liquid with 
linearized dispersions around the two Fermi points ($\propto\sqrt{h-h_c}$). 
However, with increasing $T\gtrsim h-\Delta$, the effect of band curvature becomes
more important due to thermal excitations. 

\begin{figure}[htb]
\begin{center}
\includegraphics[width=9cm]{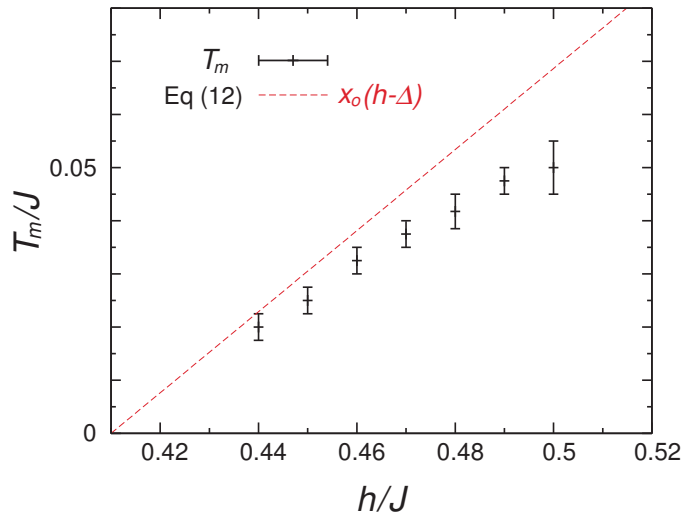}
\end{center}
\caption{The fitting of the QMC results with eq.~(\ref{eq:mini}). From
 this fitting, the gap $\Delta$ can be estimated. 
The error-bars come from a discreteness of the numerical data
with respect to temperature.}
\label{fig:mini}
\end{figure}

Here we propose a new way to estimate a gap from $T_m$. 
Equation~(\ref{eq:polylog}) takes the minimum under the condition, 
$2x={\rm Li}_{n=1/2}[-{\rm e}^x]\big/{\rm Li}_{n=-1/2}[-{\rm e}^x]$,
where $x=\mu/T_m$. Its solution, $x= x_0 \sim 0.76238$, yields 
\begin{align}\label{eq:mini}
T_m=x_0(h-\Delta).
\end{align} 
The finite temperature measurement of magnetization at several $h>h_c$ 
in both experiments and numerical calculations 
thus provide a reasonable estimation of $\Delta$. 
Since eq.~(\ref{eq:mini}) consists only of universal constant 
we can use it without any microscopic informations of the system. 
As a demonstration, we show in Fig.~\ref{fig:mini} the comparison of 
our Monte carlo results with eq.~(\ref{eq:mini}). 
The data asymptotically approaches eq.~(\ref{eq:mini}) toward $h\to
\Delta$ 
since the deviation at large $h$ is due to the interaction effect of the magnons. 
Note that all the data at $h>h_c=\Delta$ in fig.~\ref{fig:mini} are
located below the values given by 
eq.~(\ref{eq:mini}). 
This behavior is rather universal for generic models as follows;
the excitations to
the higher energy modes are enhanced by the repulsive interactions, so
that, even in low temperature regime, the band curvature effect in 
the exact results becomes more important than that in the free case. Consequently,
the exact value of the crossover temperature, $T_m$, 
must be suppressed by the repulsive interaction compared to the estimate~(\ref{eq:mini}) from the free fermion theory.

\par
\subsection{Comments on previous studies}
In light of our results, we comment on the analysis 
based on the integrability of the NL$\sigma$ model, as proposed in Ref.~\onlinecite{KonikFendley}.
They conclude that the spin-wave velocity monotonically increases 
with the increasing magnetic field,
which is contradictory with the results in the present paper. 
The reason for this discrepancy is that they introduced the finite 
magnetic field after taking the low-energy limit of the zero-field model, 
which generally does not give a correct effective theory 
above the critical field. 
To illustrate this point, 
for a moment, let us consider the $S=1/2$ Heisenberg antiferromagnetic chain as a simple example. 
The zero-field effective theory in the low-energy limit is given by the TL liquid, 
namely the free boson field theory with the Lagrangian density, 
${\cal L}_0 = \frac{1}{2} (\partial_{\mu} \phi)^2$,
where $\phi$ must be periodic, $\phi \sim \phi + 2\pi R$; R is the
compactification radius~\cite{Les}.
The application of the magnetic field after taking the the low-energy limit 
can be represented by the term,
${\cal L}_Z = - \frac{h}{\sqrt{2\pi}} \partial_x \phi.$
This formula appears to be easily handled by ``completing the square'' as 
${\cal L}_0 + {\cal L}_Z = \frac{1}{2}(\partial_{\mu} \phi')^2 -\frac{h^2}{4\pi}$, 
where $\phi' = \phi - hx/\sqrt{2\pi}$. 
Apparently, there is no renormalization of the spin-wave velocity
nor of $R$,
which is in clear contradiction to the  exact Bethe Ansatz solution~\cite{QISM}. 
Such fallacy is attributed to taking the low-energy limit first at zero-field, 
as discussed above. 
Thus one should note that although 
the irrelevant perturbations on the fixed point theory ${\cal L}_0$ can be usually 
ignored in the low-energy limit, it cannot be in deriving the effective 
theory at a finite field.
\par
The magnetization minimum has been often discussed in terms of BEC. 
Although the three-dimensional BEC observed in Ref.~\onlinecite{NOOT} 
gives similar $M-T$ curves to Fig.~\ref{fig:magh}, it differs from 
the present case in several points;  
the magnetization shows a singular cusp of minimum 
at the transition temperature, $T_c\propto (h-h_c)^{2/3}$, 
(which is however smeared in a actual systems~\cite{Jesko}), 
whereas $T_m \propto h-h_c$ in our model just marks the crossover. 
The transverse magnetization is finite, 
namely the off-diagonal long-range order is present 
in BEC at $T<T_c$, but is absent in our one-dimensional system at any temperature.

%
%
\section{Summary}
We found the characteristic temperature dependence of magnetization 
of the $S=1$ Haldane chain 
in the gapless regime at finite magnetic field, $h_c \le h \le h_s$. 
In the vicinity of the endpoints of the gapless regime, 
$h\gtrsim h_c$ and $h\lesssim h_s$, the magnetization shows 
a minimum and maximum, respectively, at the crossover temperature, $T=T_m$. 
The TL liquid theory yields the leading term of magnetization 
in the low temperature limit as, 
$M(T,h)-M(T=0,h)\propto \frac{\partial v_F(h)}{\partial h}T^2$. 
We showed that the spin wave velocity, $v_F(h)$, 
vanishes at both $h=h_c$ and $h_s$ and shows a maximum in between. 
Therefore, $M(T,h={\rm fixed})$ at around $T \sim 0$($T \le T_m$) 
undergoes a change from the decreasing function to 
the increasing function of $T$ 
as the magnetic field increases from $h=h_c$ to $h_s$. 
\par
We emphasize that the present discussions are universal  
(at least) for axial symmetric generic gapped one-dimensional spin systems, 
provided that the field-induced gapless phase can be 
described by a single-component TL liquid.
The detailed structure of $M$ and $v_F$ reflecting the characteristics 
of each system does not violate this scheme 
since we have dealt with only the universal features in the low-energy limit. 
Actually, the conclusions do not depend on a particular 
choice of the effective theory, such as the sine-Gordon model, 
the nonlinear sigma model, or the delta-function Bose gas. 
\par
We also proposed a simple and reliable scheme (represented by eq.(\ref{eq:mini})) 
to estimate the gap, $\Delta$, from the finite temperature magnetization measurement. 

To our knowledge, 
there is no reported experimental results corresponding to our analysis
in the present paper. However, it might be possible that indeed there are already some corresponding experiments, which have been interpreted in different ways.
Actually, the upturn of magnetization at $T=T_m$ 
found in many experiments had been interpreted 
as a three-dimensional effect (BEC of magnons). 
However, we argue that it does occur generally 
in purely one-dimensional gapped spin systems as well. 
Therefore, a care should be taken in the interpretation of the magnetization data.
To distinguish these two scenarios, one should examine whether or not 
the transverse magnetization exists below $T_m$ as well as 
the presence/absence of the thermodynamic phase transition.
In reality the actual magnetic systems have finite interchain interactions. 
However, the effects described in the present paper may be observable
if the interchain interactions are sufficiently weak.
We hope that the present work stimulates further study 
on one-dimensional magnetism.
\par 
\acknowledgments
We would like to thank I. Affleck, J. Akimitsu, H. Tanaka, 
C. Yasuda, K. Sakai, K. Hida, and S. Todo 
for fruitful discussions and contributions. 
This work is supported by a 21st Century COE Program 
"Nanometer-Scale Quantum Physics" at Tokyo Tech, 
a 21st Century COE Program at Aoyama Gakuin University, 
a Grant-in-Aid for Scientific Research under Grant No. 18540341 from MEXT of Japan, 
and by NSERC of Canada.

\end{document}